\documentclass{aastex62}
\usepackage{lineno}

\usepackage{soul}
\usepackage[figuresright]{rotating}
\usepackage{lineno}
\shorttitle{Timing pulsars in M3 with FAST}
\shortauthors{Li et al.}

\begin{document}

\title{Timing and Scintillation Studies of Pulsars in Globular Cluster M3 (NGC 5272) with FAST}

\author[0009-0008-4109-744X]{Baoda Li\href{https://orcid.org/0009-0008-4109-744X}{\includegraphics[scale=0.04]{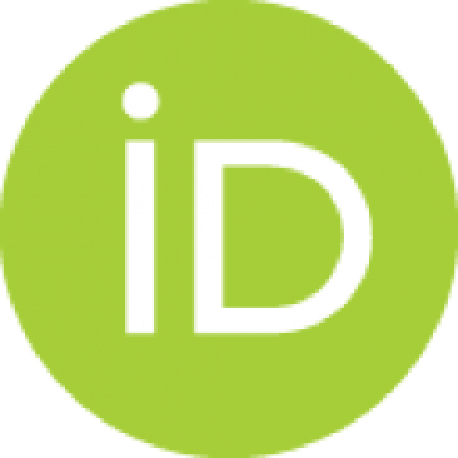}}}
\affil{College of Physics, Guizhou
University, Guiyang 550025, P.R. China}
\affiliation{International Centre of Supernovae, Yunnan Key Laboratory, Kunming 650216, P.R. China}

\correspondingauthor{Li-yun Zhang, Jumei Yao, Lei Qian, Zhichen Pan}
\email{liy\_zhang@hotmail.com}
\author[0000-0002-2394-9521]{Li-yun Zhang\href{https://orcid.org/0000-0002-2394-9521}{\includegraphics[scale=0.04]{orcid-ID.eps}}}
\affiliation{College of Physics, Guizhou
University, Guiyang 550025, P.R. China}
\affiliation{International Centre of Supernovae, Yunnan Key Laboratory, Kunming 650216, P.R. China}

\author[0000-0002-4997-045X]{Jumei Yao\href{https://orcid.org/0000-0002-4997-045X}{\includegraphics[scale=0.04]{orcid-ID.eps}}}
\email{yaojumei@xao.ac.cn}
\affiliation{Xinjiang Astronomical Observatory, Chinese Academy of Sciences, 150, Science 1-Street, Urumqi, Xinjiang 830011, P. R. China}
\affiliation{Key Laboratory of Radio Astronomy, Chinese Academy of Sciences, Urumqi, Xinjiang 830011, P. R. China}
\affiliation{Xinjiang Key Laboratory of Radio Astrophysics, 150 Science1-Street, Urumqi, Xinjiang 830011, P. R. China}

\author[0000-0001-6051-3420]{Dejiang Yin\href{https://orcid.org/0000-0001-6051-3420}{\includegraphics[scale=0.04]{orcid-ID.eps}}}
\affiliation{College of Physics, Guizhou
University, Guiyang 550025, P.R. China}

\author[0000-0001-6196-4135]{Ralph~P.~Eatough\href{https://orcid.org/0000-0001-6196-4135}{\includegraphics[scale=0.04]{orcid-ID.eps}}}
\affiliation{National Astronomical Observatories, Chinese Academy of Sciences, 20A Datun Road, Chaoyang District, Beijing 100101, P.R. China}
\affiliation{Max-Planck-Institut f\"{u}r Radioastronomie, Auf dem H\"{u}gel 69, D-53121, Bonn, Germany}

\author{Minghui Li}
\affiliation{State Key Laboratory of Public Big Data, Guizhou University, Guiyang 550025, P.R. China}

\author{Yifeng Li}
\affiliation{National Time Service Center, Chinese Academy of Sciences, Xi'an, 710600, P.R. China}
\affiliation{Key Laboratory of Time Reference and Applications,Chinese Academy of Sciences, Xi'an, 710600, P.R. China}

\author[0009-0001-6693-7555]{Yujie Lian\href{https://orcid.org/0009-0001-6693-7555}{\includegraphics[scale=0.04]{orcid-ID.eps}}}
\affiliation{Institute for Frontiers in Astronomy and Astrophysics, Beijing Normal University, Beijing 102206, P.R. China}
\affiliation{Department of Astronomy, Beijing Normal University, Beijing 100875, P.R. China}

\author{Yu Pan}
\affiliation{Chongqing University of Posts and Telecommunications, Chongqing, 40000, P.R. China}

\author[0009-0007-6396-7891]{Yinfeng.Dai\href{https://orcid.org/0009-0007-6396-7891}{\includegraphics[scale=0.04]{orcid-ID.eps}}}
\affiliation{College of Physics, Guizhou University, Guiyang 550025, P.R. China}

\author{Yaowei Li}
\affiliation{College of Physics, Guizhou
University, Guiyang 550025, P.R. China}

\author{Xingnan Zhang}
\affiliation{State Key Laboratory of Public Big Data, Guizhou University, Guiyang 550025, P.R. China}

\author{Tianhao Su}
\affiliation{College of Physics, Guizhou University, Guiyang 550025, P.R. China}

\author{Yuxiao Wu}
\affiliation{Chongqing University of Posts and Telecommunications, Chongqing, 40000, P.R. China}

\author{Tong Liu}
\affiliation{National Astronomical Observatories, Chinese Academy of Sciences, 20A Datun Road, Chaoyang District, Beijing 100101, P.R. China}

\author[0000-0002-2953-7376]{Kuo Liu\href{https://orcid.org/0000-0002-2953-7376}{\includegraphics[scale=0.04]{orcid-ID.eps}}}
\affiliation{Shanghai Astronomical Observatory, Chinese Academy of Sciences, Shanghai, 200030, P.R. China}

\author[0000-0003-0757-3584]{Lin Wang\href{https://orcid.org/0000-0003-0757-3584}{\includegraphics[scale=0.04]{orcid-ID.eps}}}
\affiliation{Kavli Institute for Astronomy and Astrophysics, Peking University, Beijing 100871, P.R. China}

\author[0000-0003-0597-0957]{Lei Qian\href{https://orcid.org/0000-0003-0597-0957}{\includegraphics[scale=0.04]{orcid-ID.eps}}}
\email{lqian@nao.cas.cn}
\affiliation{Guizhou Radio Astronomical Observatory, Guizhou University Guiyang 550025, P.R. China}
\affiliation{National Astronomical Observatories, Chinese Academy of Sciences, 20A Datun Road, Chaoyang District, Beijing 100101, P.R. China}
\affiliation{CAS Key Laboratory of FAST, National Astronomical Observatories, Chinese Academy of Sciences, Beijing 100101, P.R. China}
\affiliation{College of Astronomy and Space Sciences, University of Chinese Academy of Sciences, Beijing 100049, P.R. China}

\author[0000-0001-7771-2864]{Zhichen Pan\href{https://orcid.org/0000-0001-7771-2864}{\includegraphics[scale=0.04]{orcid-ID.eps}}}
\email{panzc@bao.ac.cn}
\affiliation{Guizhou Radio Astronomical Observatory, Guizhou University Guiyang 550025, P.R. China}
\affiliation{National Astronomical Observatories, Chinese Academy of Sciences, 20A Datun Road, Chaoyang District, Beijing 100101, P.R. China}
\affiliation{CAS Key Laboratory of FAST, National Astronomical Observatories, Chinese Academy of Sciences, Beijing 100101, P.R. China}
\affiliation{College of Astronomy and Space Sciences, University of Chinese Academy of Sciences, Beijing 100049, P.R. China}

\begin{abstract}
We present the phase-connected timing solutions of all the five pulsars in globular cluster (GC) M3 (NGC~5272), namely PSRs~M3A to F (PSRs~J1342+2822A to F),
with the exception of PSR~M3C, from FAST archival data.
In these timing solutions,
those of PSRs~M3E, and F are obtained for the first time.
We find that PSRs~M3E and F have low mass companions, and are in circular orbits with periods of 7.1 and 3.0~days, respectively.
For PSR~M3C, we have not detected it in all the 41 observations.
We found no X-ray counterparts for these pulsars in
archival Chandra images in the band of 0.2–-20~keV.
We noticed that the pulsars in M3 seem to be native.
From the Auto-Correlation Function (ACF) analysis of the M3A's and M3B's dynamic spectra, the scintillation timescale ranges from $7.0\pm0.3$\,min to $60.0\pm0.6$\,min, and the scintillation bandwidth ranges from $4.6\pm0.2$\,MHz to $57.1\pm1.1$\,MHz. The measured scintillation bandwidths from the dynamic spectra indicate strong scintillation, and the scattering medium is anisotropic. From the secondary spectra, we captured a scintillation arc only for PSR~M3B with a curvature of $649\pm23\,{\rm m}^{-1}\,{\rm mHz}^{-2}$.

\end{abstract}

\section{Introduction}\label{sec:intro}

Globular clusters (GCs) with very high stellar number density have been proven to be factories of
millisecond pulsars (MSPs) \citep{2017MNRAS.471..857F}.
The GC M3 (NGC~5272) is located 10.2\,kpc away \citep{1996AJ....112.1487H}.
At present, six pulsars have been found in this cluster,
namely PSRs~M3A to F, or PSRs~J1342$+$2822A to F.
All of them are MSP binaries with dispersion measures (DMs) around $26.4\,{\rm pc}\,{\rm cm}^{-3}$.
PSRs~M3A to D were found by Arecibo and Green Bank Telescope (GBT) \citep{2005ASPC..328..199R,2007ApJ...670..363H},
while PSRs~M3E and F \citep{2021ApJ...915L..28P} were found by the Five-hundred-meter Aperture Spherical radio Telescope \citep[FAST, ][]{2011IJMPD..20..989N}.

PSR~M3A is a black widow with a minimum companion mass of 0.01\,M$_{\odot}$ and an orbital period of 0.14\,days \citep{2021ApJ...915L..28P}.
PSR~M3B has a minimum companion mass of 0.21\,M$_{\odot}$ and an orbital period of 34.0\,hours \citep{2007ApJ...670..363H}.
PSR~M3C was detected only once by GBT~\citep{2005ASPC..328..199R}, and is still to be confirmed at other telescopes.
PSR~M3D has a companion with a minimum mass of 0.21\,M$_{\odot}$ and an orbital period of 129\,days.
Such an orbit indicates that PSR~M3D probably experienced abnormal evolution \citep{2007ApJ...670..363H}.

Due to scintillation, the detection rates of pulsars in M3 from previous studies are quite low.
For example, in 78 observations with Arecibo and GBT,
PSRs~M3A to D only appeared three times, 16 times, once (at GBT), and 12 times, respectively \citep{2007ApJ...670..363H}.
Only the timing solutions of PSRs~M3B and D have been obtained.

When a pulsar radio signal propagates through the turbulent ionized interstellar medium (ISM), the relative motion between the pulsar, the ISM and the Earth lead to scintillation,
which changes the pulsar intensity \citep{1990ARA&A..28..561R}.
There are two main types of pulsar scintillation, namely refractive interstellar scintillation (RISS) and diffractive interstellar scintillation (DISS).
RISS from the large-scale inhomogeneity of the ISM leads to a long-term (e.g., days to months) fluctuation of pulsar flux densities \citep{1984A&A...134..390R,1986ApJ...301L..53B}. DISS from the small-scale inhomogeneities in the ISM leads to short time-scale (e.g., minutes to hours) variations of pulsar flux densities \citep{1969Natur.221..158R}.

For DISS, scintillation bandwidth $\Delta\nu_{\rm d}$ and scintillation timescale $\Delta\tau_{\rm d}$ can be obtained from the Auto-Correlation Function (ACF) analysis of a pulsar dynamic spectrum,
which is a map of pulsar intensity as a function of observing frequency and time.
The $\Delta\tau_{\rm d}$ is the half width of ACF at $1/e$ along the time axis,
and $\Delta\nu_{\rm d}$ is the half width at the half maximum along the frequency axis \citep{1986ApJ...310..737C}.
The scintillation bandwidth can be used to determine the scintillation strength \citep{1990ARA&A..28..561R}. Studies on pulsar scintillation can help us to reveal properties of the ISM, for example, its anisotropic or isotropic nature \citep{2005MNRAS.358..270W}. The secondary spectrum is the squared modulus of the two dimensional Fourier transform of a dynamic spectrum.
In some situations, we can detect so-called scintillation arcs (one or many) in the secondary spectrum \citep{2001ApJ...549L..97S}.
For pulsars with distance and proper motion measurements,
it is possible to obtain the density, scale and location of screen-like ISM structures through the detection of the scintillation arcs or arclets in the secondary spectrum \citep{2005ApJ...619L.171H,2020RAA....20...76Y}.
For pulsars in binary systems,
the scintillation timescale and the arc curvature are modulated by orbital motion,
and their measurements may enable us to constrain both the binary orbital inclination and the distance of the scattering screen \citep{2019MNRAS.485.4389R}.

With FAST, the detection rates of pulsars in M3 are more than twice, those seen
in previous studies.
This increased rate provides a better opportunity to study the properties of pulsars in M3 though both pulsar timing and observations of scintillation along the line of sight to this GC.
In this paper, we update the ephemerides of PSRs~M3A, B, and D, and report the timing solutions of PSRs~M3E and F and an analysis of the scintillation properties of PSRs~M3A and B in M3.
In Section~\ref{sec:data} the data reduction methods used are described.
Section~\ref{sec:dis} presents
the results and discusses the detection rates and timing solutions; encounter rates and eccentricities; cluster acceleration; and  scintillation properties. In Section~\ref{sec:sum}, we give a short summary.

\section{Data Reduction}\label{sec:data}
We analyzed the FAST archival data of all the 41 observations.
All these data were recorded with the 19-beam L-band receiver covering a frequency range of 1.0 to 1.5\,GHz \citep{2020RAA....20...64J}.
Because the core radius of M3 ($0.37'$) fits within the half-power beamwidth of a single receiver ($3'$), only data from the central beam was used in our study.
The signals were 8-bit sampled with two or four polarization products and recorded in {\sc psrfits} format \citep{2004PASA...21..302H}.
The number of frequency channels used is 4096 (corresponding to a 0.122\,MHz channel width),
and the sampling time is 49.152\,{\rm $\mu$s}.
All 41 observations are pointed toward the center of M3 at right ascension (RA)~(J2000) 13:42:12.00 and declination (DEC)~(J2000) +28:22:38.2,
with observation times range from 20\,minutes to 5\,hours on dates from MJD 58831 to 59767.

\subsection{Data Reduction for Timing}
According to the previous studies,
PSRs~M3B (orbital period $\sim$1.4~day) and D ($\sim$129~day) have timing solutions \citep{2007ApJ...670..363H},
PSR~M3A was proved to be a black widow ($\sim$ 0.14~day) based on only two observations \citep{2021ApJ...915L..28P}.
The published ephemerides of PSRs~M3B and D have not been updated for almost 20 years, and  cover MJDs~52763 to 53542 and 52768 to 53476, respectively \citep{2007ApJ...670..363H}.
PSR~M3C, if it is a real pulsar, was only seen once by GBT \citep{2005ASPC..328..199R} but is believed to be in a binary system \citep{2007ApJ...670..363H}.
In order to re-detect all the pulsars in M3, we folded the data with the timing solutions, and also performed blind searches to account for any significant offsets accrued in the spin frequency over the intervening years, or from unknown orbital modulation.
The software used for searching and folding was PRESTO \footnote{
https://github.com/scottransom/presto} \citep{2001PhDT.......123R,2011ascl.soft07017R} and TEMPO\footnote{http://pulsar.princeton.edu/tempo} was used to either create or update timing solutions. The routine {\sc rfifind} from PRESTO was used for generating radio frequency interference (RFI) masks.
The length of data for finding radio frequency domain RFI ({\tt -time} option) was 12.9\,s.
The routine {\sc prepdata} was used to dedisperse the data.
Considering the possible range of DMs in a given GC \citep{2023RAA....23e5012Y},
the range used for M3 was 24.5 to 28.5\,pc\,cm$^{-3}$.
In order to catch any possible faint signals, a DM step of 0.01 pc~cm$^{-3}$ was used for dedispersion.
Then, a Fast Fourier Transform was applied to the dedispersed time series to obtain fluctuation spectra.
The routine {\sc accelsearch} was used to search for possible periodic signals in the spectra where the largest range in acceleration (defined by the {\sc accelsearch} parameter {\tt -zmax}) was set to be 1200, where {\tt -zmax} represents the number of Fourier bins that a signal can drift during an individual observation.
All periodic signals with a clearly defined DM
were folded using the routine {\sc prepfold} to produce so-called candidate plots that were visually checked.
As a result, PSRs~M3A, B, D, E, and F were detected on multiple occasions (24, 37, 32, 21 and 33 times, respectively) enough to construct updated timing solutions.

\begin{figure*}
\begin{center}
\includegraphics[width=18cm,height=10cm]{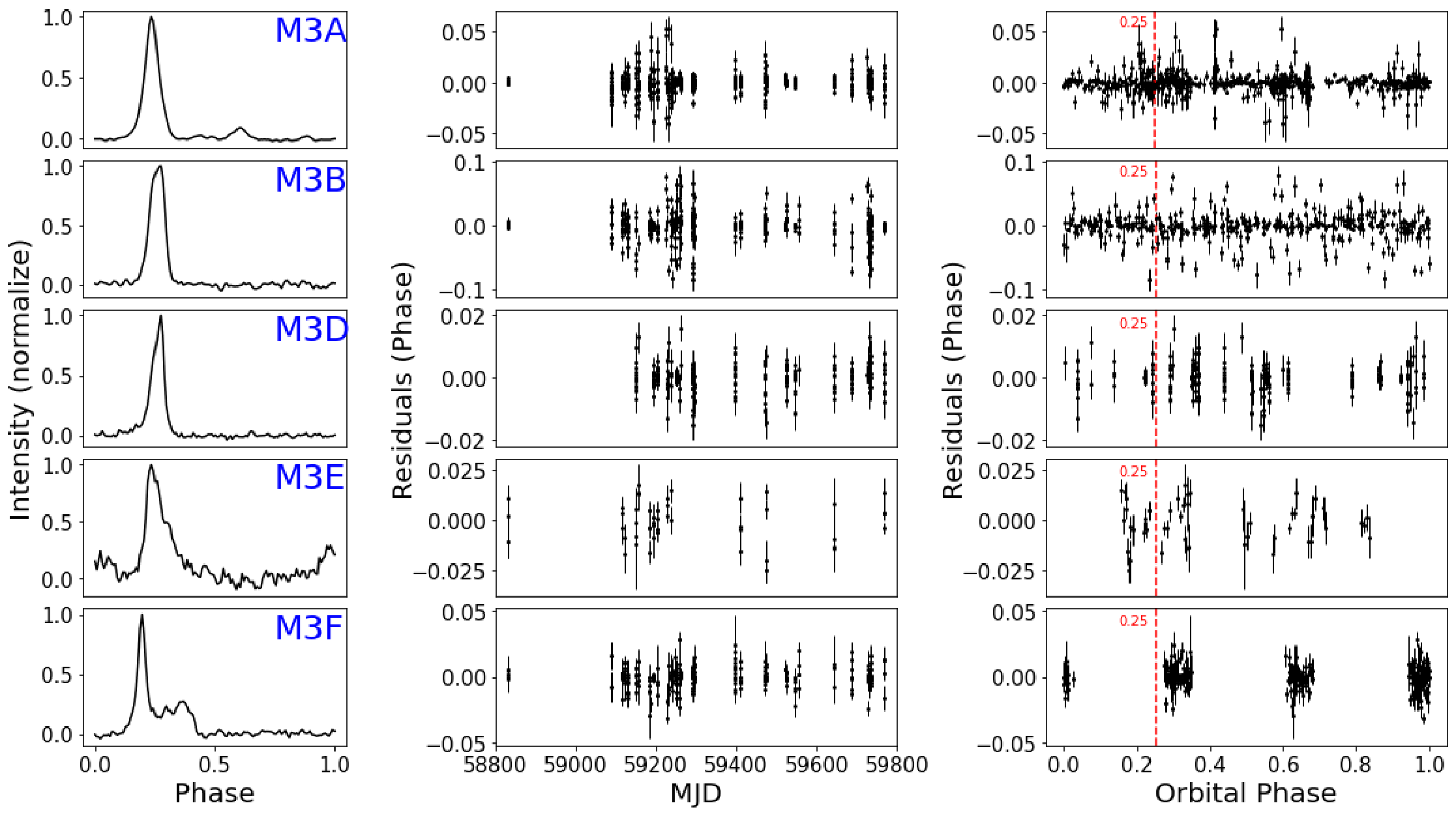}
\caption{Profiles of PSRs~M3A, B, D, E, and F (left panels), timing residuals for PSRs~M3A, B, D, E, and F as a function of MJD (middle panels), and orbital phase (right panels).
In the right panels, the orbital phase of 0.25 is marked by the red dashed lines, which is superior conjunction for pulsars in circular orbits.
Except for PSR~M3F, which was not covered, none of them showed eclipsing phenomena.}\label{fig:1}
\end{center}
\end{figure*}

The routine {\sc get\_TOAs.py} from PRESTO was used to obtain the times of arrivals (ToAs).
Standard reference profile templates for correlation (to measure ToAs) were obtained by fitting the pulse profiles from the detection with the highest signal-to-noise ratios (SNRs) with the PRESTO routine {\sc pygaussfit.py}.
According to the SNR of the pulsar singal, the number of ToAs obtained from each observation ranged from 4 to 16.
For PSRs~M3E and F,
we used the {\sc fitorb.py} routine\footnote{https://github.com/scottransom/presto/blob/master/bin/fitorb.py} to fit for the initial orbital parameters.
With the initial parameter files from either the {\sc psrcat} (PSRs~M3A, B, and D) or orbit fitting (PSRs~M3E and F),
the phase-connected timing solutions for PSRs~M3A, B, D, E, and F were obtained.
Figure~\ref{fig:1} shows integrated pulse profiles and timing residuals from the FAST observations and Table~\ref{tab:1} presents the timing solutions for the five M3 pulsars.
\begin{sidewaystable}[thp]
\begin{center}
\caption{{\bf Timing solutions of PSRs~M3A, B, D and PSRs~M3E, F. The Solar System Ephemeris we used is DE438, Time Units is TDB, and we did not use the clock correction file.}}\label{tab:1}
\resizebox{\textwidth}{90mm}{
\setlength{\tabcolsep}{1mm}{
\renewcommand{\arraystretch}{1.5}
\begin{tabular}{l c c c c c }
\hline
Pulsar                                    &M3A      &    M3B      &   M3D   &    M3E   &    M3F     \\
\hline\hline
Right Ascension, $\alpha$ (J2000)\dotfill & 13:42:13.79168(8)  &13:42:11.08686(2)     &13:42:11.35220(9)  &13:42:12.3908(4)  &13:42:12.37908(6)                    \\
Declination, $\delta$ (J2000)\dotfill     & 28:23:01.966(1)    &28:22:40.0844(2)       &28:22:30.160(1) &28:23:16.131(6)   &28:22:37.270(1)                    \\
Spin Frequency, $f$ (s$^{-1}$)\dotfill    &392.96772541864(1)  &418.5114725920(2) &183.7230520500(1) &182.7196423300(2) &227.07633075643(2)  \\
1st Spin Frequency derivative, $\dot{f}$ (Hz s$^{-2}$)\dotfill & 1.5704(8)$\times 10^{-15}$ &$-$3.2743(3)$\times 10^{-15}$  &   $-$9.47(2)$\times 10^{-16}$  &3.5(7)$\times 10^{-17}$ &$-$1.610(1)$\times 10^{-15}$ \\
Reference Epoch (MJD)\dotfill             &59194.14       &52770.00      &52770.00 &59203.93     & 59291.67     \\
Start of Timing Data (MJD)       \dotfill &58830.98          &58830.98         &59149.08  &58831.00         &58830.99       \\
End of Timing Data (MJD)\dotfill          &59767.57          &59767.58         &59767.52  &59767.57         &59767.58       \\
Dispersion Measure, DM (pc cm$^{-3}$)\dotfill &26.4278(8)      &26.1510(4)         &26.3677(3)  &26.521(7)     &26.436(2)        \\
Number of TOAs                   \dotfill &388                &585               &239 &71                &254              \\
Residuals RMS ($\mu$s)           \dotfill &20.33              &7.11              &12.05  &36.27              &19.66           \\
\hline\hline
Binary Model\dotfill                                  &   ELL1            &ELL1              &   BT  &ELL1        &ELL1              \\
Projected Semi-major Axis, $x_{\rm p}$ (lt-s)\dotfill &0.0256212(7)    &1.8756549(3)   &38.521421(2)&5.33923(1)     &2.225893(4)     \\
Orbital Eccentricity, $e$    \dotfill                 & -- &-- &7.47556(1)$\times 10^{-2}$ &-- &--\\
Longitude of Periastron, $\omega$ (deg)\dotfill       &--   &--                &206.18223(2) &--        &--    \\
Epoch of passage at Periastron, $T_0$ (MJD) \dotfill  & -- &--    &52655.44931(5) &--  &--\\
First Laplace-Lagrange parameter, $\eta$ \dotfill     & 3(5)$\times 10^{-5}$           &4.0(3)$\times 10^{-6}$  &   --  &3.04(5)$\times 10^{-4}$ &3(4)$\times 10^{-6}$  \\
Second Laplace-Lagrange parameter, $\kappa$  \dotfill  & $-$6(5)$\times 10^{-5}$ &6(4)$\times 10^{-7}$  &  -- &$-$5.5(4)$\times 10^{-5}$   &4(5)$\times 10^{-6}$  \\
Epoch of passage at Ascending Node, $T_\textrm{asc}$ (MJD) \dotfill &58831.220370(2) &52485.967961(2) & --&59195.297476(4)&59231.715533(1)\\
Orbital Period, $P_b$ (days)\dotfill     &0.1358561044(3)       &1.4173522991(3) &128.7454866(5)  &7.0968527(1)  &2.991987692(4)  \\
\hline\hline
Offset from GC center in $\alpha$, $\theta_\alpha$\dotfill & -0.4775 & -0.1173 & +0.0589 & -0.1695 & +0.1670 \\
Offset from GC center in $\delta$, $\theta_\delta$\dotfill & +0.3961  & +0.0314 & -0.1340 & +0.6322 & -0.0155 \\
Total offset from GC center, $\theta_\perp$ (arcmin)\dotfill & 0.6204 & 0.1214 & 0.1464 & 0.6545 & 0.1677 \\
Total offset from GC center, $\theta_\perp$ (core radii)\dotfill &1.6768 & 0.3281 & 0.3956 & 1.7690 & 0.4532 \\
Projected distance from GC center, $r_\perp$ (pc)\dotfill & 1.8408 & 0.3602 & 0.4343  & 1.9420 & 0.4975 \\
 \hline
\end{tabular} }}
\end{center}
\end{sidewaystable}

\subsection{Data Reduction for Scintillation}
{\sc prepfold} from PRESTO was used to create the standard period folded archive files that are the basis for scintillation studies.
Each archive had a sub-integration time and frequency channel resolution of 60~s and 0.98~MHz, respectively.
Then, {\sc pazi}, {\sc pam}, and {\sc psrflux } from PSRCHIVE were used to analyze the data \citep{2012AR&T....9..237V}.
Then, {\sc pazi} was used to manually remove RFI.
Second, the routine {\sc pam} was used to sum the polarization channels to total intensity.
Third, {\sc psrflux} was used to generate the dynamic spectrum from the archive files.

The scintillation bandwidth and scintillation timescale can be obtained from the ACF analysis of the dynamic spectrum.
The ACF $C(\Delta\nu,\Delta\tau)$, where $\Delta\nu$ and $\Delta\tau$ are the frequency and time lag, respectively, is defined by the formula (\citealt{2023SCPMA..6699511Z} and references therein): 
\begin{eqnarray}
C(\Delta\nu,\Delta\tau) = \sum_{\nu}{}\sum_{t}{} \Delta D(\nu, t) \Delta D(\nu + \Delta\nu, t + \Delta\tau),
\end{eqnarray}
where $D(\nu, t)$ is the pulsar intensity as a function of frequency, $\nu$ and time, $t$.
 Here $\Delta D(\nu, t)$ = $D(\nu, t) - \overline{D}$ (where ``${-}$" denotes averaging).
The normalized ACF is $N(\Delta\nu,\Delta\tau) = C(\Delta\nu,\Delta\tau)/C(0,0)$.
Following \citet{2020ascl.soft11019R},
we use their Scintools package\footnote{https://github.com/danielreardon/scintools} to calculate the ACF in this work.

The secondary spectrum is obtained from the two dimensional Fourier transform of the dynamic spectrum.
We got the secondary spectrum with the equation \citep{2020ascl.soft11019R}
\begin{eqnarray}
P(f_{t},f_{\lambda}) =10\log_{10}(\mid\tilde{D}(t,\lambda)\mid^{2}),
\end{eqnarray}
where the $\tilde{ }$ symbol denotes the two-dimensional Fourier transform,
and $f_{t}$ and $f_{\lambda}$ are the conjugate time and the conjugate wavelength, respectively.
The transformed data are squared, shifted and cropped for the study.

\section{Results \& Discussion}\label{sec:dis}

\begin{figure*}
\begin{center}
\includegraphics[width=0.721\linewidth]{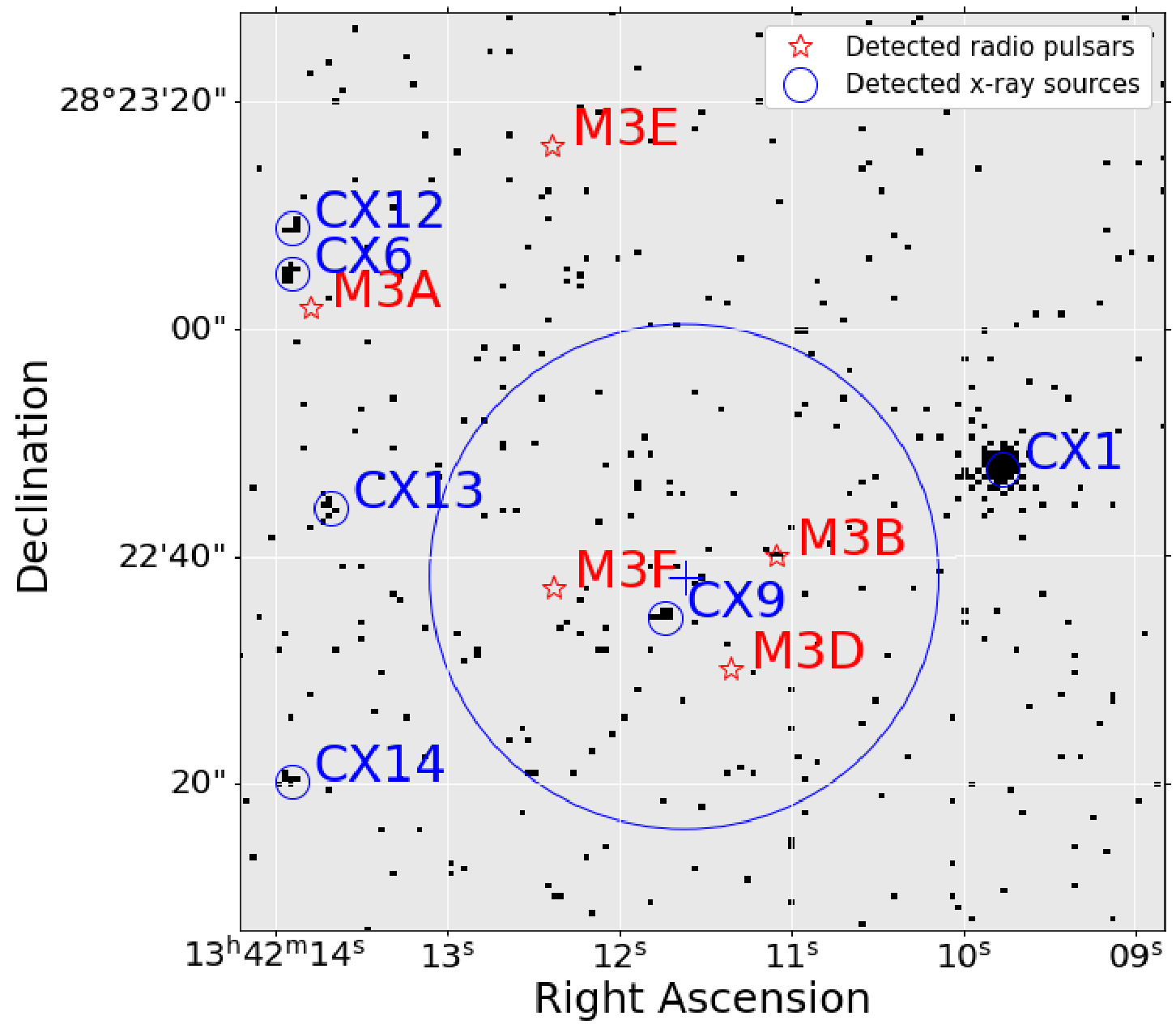}
\caption{The radio pulsars and the X-ray sources of M3 in the Chandra image. PSRs M3A, B, D, E, and F are presented by the red stars.
The X-ray sources are marked by the small blue circles.
The center of M3 (J2000, RA 13:42:11.62 DEC +28:22:38.2) is shown as a blue cross.
The large blue circle represents the core radius of 0.37~arcminutes.
The energy range for the background X-ray image is from 0.2 to 20~keV.}\label{fig:2}
\end{center}
\end{figure*}

\subsection{Detection Rates and Timing Solutions} \label{sec:Timing result}
The known pulsars in M3 are all in binary systems and
all show scintillation.
This can be seen from the fluctuations in the detected folded SNR given in Table \ref{tab:2}.
The previous searches for pulsars in M3 were carried out with Arecibo \citep{2007ApJ...670..363H} where observations covered a frequency range of 1.1 to 1.6~GHz.
The corresponding detection rates of PSRs~M3A to D were 4\%, 21\%, once, and 15\%, respectively.
With FAST, the detection rates of pulsars in M3 are greatly improved, except for PSR M3C.
In the 41 FAST archival data,
PSRs M3A to F were detected for 24 (58\%), 37 (90\%), 0, 32 (78\%), 21 (51\%), and 33 (80\%) times, respectively (see Table \ref{tab:Detection Rates}).

In order to catch PSR~M3C,
we searched for it with the whole length of the data and some 5, 30, and 100~minutes segments.
In all these searches, the value of the {\sc accelsearch} option which determines the range of accelerations searched ({\tt -zmax}) was set to 1200.
Then, we searched for any possible signals which have periods around 2.166~ms (PSR~M3C).
Unfortunately, no pulsar-like signals were detected.
Therefore, we postulate that PSR~M3C was discovered undergoing extreme scintillation -- like PSR~M30B in \citet{2004ApJ...604..328R} -- which has recently been detected again with MeerKat only for the second time in \citet{2023ApJ...942L..35B}, or the detection candidate was possibly due to noise fluctuation.

With the FAST data,
we derived the phase-connected timing solutions of PSRs M3A, B, D, E, and F.
The results of PSRs M3A, B, and D were consistent with the previous studies: PSR~M3A~\citep{2021ApJ...915L..28P} and PSRs~M3B~and~D~\citep{2007ApJ...670..363H},
with slight differences in positions, spin frequencies, and spin frequency derivatives.
For example, the DEC, $\delta$ of PSR~M3B is {+28:22:40.141(2)} in \citet{2007ApJ...670..363H}, and the result given here is {+28:22:40.0844(2)}; a small difference of $0.06''$.
The difference in $\delta$ is a slight difference may be caused by the longer baseline and more timing data (16 detections covering 779 days in \citet{2007ApJ...670..363H}, 37 detections covering 937 days in this work).

PSR M3A is a black widow in a 0.14-day orbit.
The minimum companion mass is around 0.01\,M$_{\odot}$.
No eclipsing was observed.
The timing solution of PSR~M3D was greatly improved with errors in RA and DEC being reduced by factors of 1000 and 10,000 times, respectively.

PSRs~M3E~and~F are two pulsars recently discovered by FAST \citep{2021ApJ...915L..28P}.
With a detection rate of ~50\%,
PSR M3E has a period of 5.473\,ms and a DM value of 26.5\,pc cm$^{-3}$.
It is in an orbit of 7.1~days with a minimum companion mass of 0.2\,M$_{\odot}$.
The eccentricity of its orbit is small but detectable, being $\rm \sim$4.5$\times$10$^{-4}$.
Showing faint signals,
PSR M3F has a high detection rate of 80\%, only 10\% lower than that of PSR M3B.
Its period and DM are 4.404\,ms and 26.4\,pc cm$^{-3}$, respectively.
It is in an orbit of 3~days with a minimum companion mass of 0.17\,M$_{\odot}$.
It seems that all the 33 detections of PSR~M3F only appeared on the orbital phases of approximately 0.00, 0.33, and 0.67.
The orbital period of PSR M3F is almost exactly 3 sidereal days in duration.
And since FAST is (mostly) a transit telescope, when we observe on a random day over a relatively short timespan (a year or two), we
sample only 1 of 3 different orbital phases.
It will take a long time for the very small deviation of 
$P_{\rm b}/3$
from a sidereal day to allow \st{full }sampling of the full orbit.

\begin{table}[]
\centering
\caption{The SNRs for all observations of M3 pulsars (except for PSR ~M3C) made at FAST.
For each pulsar, the lowest and the highest SNRs are marked with wavy line and underline, respectively.\label{tab:2}}
\setlength{\tabcolsep}{5.0mm}{
\begin{tabular}{|c|c|c|c|c|c|c|c|}
\hline
Data     & MJD   & Observation & M3A & M3B & M3D & M3E & M3F   \\ 
YYYYMMDD &       & Length (s)  & (SNR) & (SNR) & (SNR) & (SNR) & (SNR)   \\ \hline
20191214 & 58831 & 16200 & 28.2  & 67.3  &  & 4.0  & 7.8   \\ \hline
20200828 & 59089 & 1800  &  & 10.4  &  &  &   \\ \hline
20200923 & 59115 & 7200  & 9.8  & 41.4  &  & 5.0  & 15.8   \\ \hline
20200929 & 59121 & 7200  &  & 10.5  &  & {\uwave{2.9}}  & 7.0    \\ \hline
20201008 & 59130 & 7200  & 6.0  & 16.4 & 5.4  &  & 6.3    \\ \hline
20201027 & 59149 & 14400 &  & {\uwave{4.9}}  & 26.4  & 3.8  & 8.7   \\ \hline
20201102 & 59155 & 18000 &  & 10.3  & 4.6  & 4.0  &   \\ \hline
20201130 & 59183 & 18000 & 13.7  & 18.0  & {\uwave{3.5}}  & 5.9  & 5.5   \\ \hline
20201205 & 59188 & 5880  & 10.2  &  &  &  &   \\ \hline
20201211 & 59194 & 18000 & 34.2  & 16.7  & 30.0  & 6.3  & 23.8   \\ \hline
20201221 & 59204 & 11115 & 8.4  & 7.5  & 22.7  & 5.6  &   \\ \hline
20210112 & 59226 & 18015 &  & 29.8  & 11.9  & 7.6  &  \\ \hline
20210117 & 59231 & 2715  &   &  & 7.7  &  &  \\ \hline
20210125 & 59239 & 18015 & 4.7  & 17.7  & 8.6  & 4.8  & {\underline{24.1}}   \\ \hline
20210126 & 59240 & 18015 & 18.0  & 50.7  & 5.3  & 3.6  & {\uwave{4.8}}   \\ \hline
20210204 & 59249 & 18015 &  & 33.3  & 7.6  &  & 10.2  \\ \hline
20210205 & 59250 & 14160 & 4.2  & 69.2  & 41.6  & 4.2  & 10.2   \\ \hline
20210214 & 59259 & 14400 &  & 22.7  & 8.7  &  &   \\ \hline
20210215 & 59260 & 18015 & {\uwave{4.0}} & 18.0  & 4.0  &  & 11.9   \\ \hline
20210318 & 59291 & 18015 & 6.6  & 16.3  & 6.6  &  & 10.0   \\ \hline
20210319 & 59292 & 18015 &  & 37.2  & 10.0  &  & 17.0   \\ \hline
20210320 & 59293 & 18015 & 6.2  & 51.1  & 27.0  & 3.6  & 6.2   \\ \hline
20210321 & 59294 & 19184 &  & 63.4  & 7.9  & 5.1  & 10.6   \\ \hline
20210701 & 59396 & 18000 & 11.0  & 14.8  & 10.0  &  & 8.4  \\ \hline
20210702 & 59397 & 18320 & 4.7  & 69.1  & 8.0  &  & 6.8  \\ \hline
20210714 & 59409 & 16377 & 7.1  & 9.7  &  & 6.0  & 8.1   \\ \hline
20210914 & 59471 & 14400 & 4.5  & 10.5  & 16.6  & 3.6  & 9.6  \\ \hline
20210915 & 59472 & 14400 & 13.4  &  & 6.2  &  & 23.3  \\ \hline
20210916 & 59473 & 14400 &  & 36.6  & 9.0  & 5.6  & 18.7   \\ \hline
20211105 & 59523 & 14400 & 22.2  & 33.0  &  &  &   \\ \hline
20211106 & 59524 & 11400 &  & {\underline{70.5}}  & 24.4  &  & 9.1   \\ \hline
20211127 & 59545 & 13800 & 32.8  & 32.6  & 15.4  & 5.4  & 5.7   \\ \hline
20211208 & 59556 & 14400 &  &  &  &  & 14.6   \\ \hline
20220307 & 59645 & 11115 & 5.8  & 30.8  & 13.2  & 5.1  & 4.1   \\ \hline
20220418 & 59687 & 5040  &  & 22.9  & 21.9  & 4.0  & 7.1   \\ \hline
20220527 & 59726 & 7980  &  & 23.0  & {\underline{45.4}}  &  &   \\ \hline
20220530 & 59729 & 14715 & 11.9  & 35.1  & 5.0  &  & 16.0   \\ \hline
20220601 & 59731 & 7740  &  & 40.5  & 8.4  &  & 13.2   \\ \hline
20220604 & 59734 & 4380  &  & 40.8  & 6.8  &  & 10.6   \\ \hline
20220606 & 59736 & 5880  & 8.9  & 14.1  &  &  & 9.9   \\ \hline
20220707 & 59767 & 16980 & {\underline{46.2}}  & 10.2  & 13.9  & {\underline{7.8}}  & 6.0   \\ \hline
\end{tabular}}
\end{table}

\begin{table}[]
\centering
\caption{Comparison of M3 pulsar detection rates (given as a percentage of observations performed) at three telescopes. For brevity we have removed the ``PSR'' notation from each pulsar name and ``-" indicates either un-published or non-applicable detection rates.\label{tab:Detection Rates}}
\begin{tabular}{c|cccccc}
& M3A & M3B & M3C & M3D & M3E & M3F \\
\hline\hline
FAST & \multicolumn{1}{r}{58\%} & \multicolumn{1}{r}{90\%} & - & \multicolumn{1}{r}{78\%} & \multicolumn{1}{r}{51\%} & \multicolumn{1}{r}{80\%} \\
Arecibo & \multicolumn{1}{r}{4\%} & \multicolumn{1}{r}{21\%} & - & \multicolumn{1}{r}{13\%} & - & - \\
GBT & - & - & \multicolumn{1}{r}{once} & - & - & - \\
\hline
\end{tabular}
\end{table}

\subsection{X-ray source associations}\label{sec:xray}
The core radius of M3 is $0.37'$.
The PSRs M3A and E are out of the core region while PSRs M3B, D, and F are in the core region.
Generally, rotation-powered MSPs are weak X-ray sources \citep{2021ApJ...912..124B}.
Fig. \ref{fig:2} shows the position of M3 pulsars overlayed on an X-ray image from archival Chandra data.
Six bright X-ray sources \citep{2019MNRAS.483.4560Z} near the GC center were detected by Chandra (see blue circles in Fig. \ref{fig:2}).
The known pulsars are represented by the red stars.
No X-ray counterparts were identified for any pulsar in M3.

\subsection{The Single-Binary Encounter Rate and Eccentricities}

The percentage of binary pulsars in a GC is related to the single-binary encounter rate \citep{2023ApJ...951L..37L},
\begin{equation}
\gamma \propto \frac{\rho_{\rm c}}{\upsilon},
\end{equation}
where $\rho_{\rm c}$ is the GC central density and
$\upsilon$ is the velocity dispersion \citep{2023ApJ...951L..37L}.
When $\gamma$ is normalized to the value for M4,
being $\gamma_{\rm M4}$ as in \citet{2014A&A...561A..11V}.
The $\gamma_{{\rm M4}}$ of M3 is 0.62.
The pulsars in M3 are all in binary systems, it is similar to other GCs with low $\gamma_{{\rm M4}}$, e.g. M13, $\gamma_{{\rm M4}}$ = 0.52, four binaries and two isolated \citep{2020ApJ...892...43W}.
In GCs with small $\gamma_{{\rm M4}}$, X-ray binaries are expected to evolve without much interference, forming binary MSPs similar to those in the Galactic disk.

In M3, the eccentricity of pulsars is close to 0, except for PSR~M3D (0.075).
It is in a very long orbit of about 129~days.
Its orbital period is much longer than the typical orbital period (1~day) of GC MSPs.
Only 16 among 177 GC binary pulsars\footnote{https://www3.mpifr-bonn.mpg.de/staff/pfreire/GCpsr.html (accessed on 15 January 2024)} are with orbital periods larger than 10~days.
They are believed to be formed through an exchange interaction between an isolated neutron stars and primordial cluster binaries,
and pulsars with orbital periods greater than 100 days tend to reside in low-density clusters with the central density $\rho_{\rm c}$ less than 4~log$_{10}$L$_{\odot}$~pc$^{-3}$ \citep{1992PASP..104..981H,1993ApJ...415..631S}.
As M3 has a central density $\rho_{\rm c}$ = 3.72~log$_{10}$L$_{\odot}$~pc$^{-3}$, PSR M3D follows this trend.
Its eccentricity is likely the result of passing-by encounters with other binaries.
The time scale needed to produce the eccentricity of PSR M3D can be estimated as in \citet{1995ApJ...445L.133R,2023ApJ...951L..37L}:
\begin{eqnarray}
t_{e} \simeq 4 \times 10^{11} {\rm yr } (\frac{n}{10^{4}{\rm pc}^{-3}})^{-1}(\frac{v}{10{\rm km s}^{-1}}) \times (\frac{P_{b}}{{\rm days}})^{-2/3} e^{2/5},
\end{eqnarray}
where $n$ is the number density of stars ($n \propto \rho_{c}$,
$\rho_{\rm c}$ is the central density of GC,
which is 3.72 $\times$ 10$^{3}$~L$_{\odot}$~pc$^{-3}$ for M3 \citep{1996AJ....112.1487H}),
$v$ is the one-dimensional core velocity dispersion,
which is 7.8~km~s$^{-1}$ for M3\footnote{https://people.smp.uq.edu.au/HolgerBaumgardt/globular/orbits.html} \citep{2022ApJ...934..150L},
$P_{\rm b}$ is the orbital period,
and $e$ is the eccentricity.
Following \citet{2023ApJ...951L..37L},
$n$ is roughly estimated as 8.9 $\times$ 10$^{3}$~pc$^{-3}$ for M3.
The estimated $t_{e}$ is $\thicksim$ 4.8~Gyr,
and the characteristic age of M3 is 8.7~Gyr \citep{2021JAVSO..49..192H}.
This shows that PSR M3D has had enough time to accumulate the eccentricity caused by disturbance.


\subsection{Acceleration Caused by Cluster Gravity}\label{sec:Acceleration}
From timing results, we noticed that the spin period derivatives of PSRs M3A and E are negative.
Usually, the intrinsic spin period derivative of a pulsar should be positive.
Like other pulsars in GCs, the negative spin period derivative is
expected to be caused by
the contribution of the gravitational potential of the GC \citep{1993ASPC...50..141P}.
In general, the observed spin period derivative can be described by
\begin{eqnarray}
c(\frac{\dot{P}}{P})_{{\rm obs}} - c(\frac{\dot{P}}{P})_{{\rm int}} = \frac{V_{\bot}^{2}}{d} + a_{\rm g} +a_{\ell,GC}, \label{eq:4}
\end{eqnarray}
where
$P$ is the pulsar spin period, $\dot{P}$ is the spin period derivative,
$V_{\bot}$ is the transverse velocity of the system,
$d$ is the distance to the pulsar,
$a_g$ is the acceleration due to the Galactic potential, $a_{l,GC}$ the line-of-sight acceleration due to the GC and
$c$ is the speed of light.
The term $\frac{V_{\bot}^{2}}{d}$ is the contribution of the Shklovskii effect \citep{1970SvA....13..562S}.
The proper motion and the distance of pulsars in M3 have not been determined, however, they should be similar to that of M3 itself.
    The proper motion of M3 in RA~$\alpha$ and DEC~$\delta$
is -0.153$\pm$0.007~mas~yr$^{-1}$ and -2.679$\pm$0.007~mas~yr$^{-1}$, respectively.
The distance, $d$, is $\thicksim$ 10.2~kpc \citep{2021MNRAS.505.5978V}.
So, we can get $\frac{V_{\bot}^{2}}{d}$ $\thicksim$ 5.349 $\times$ 10$^{-11}$~m~s$^{-2}$.
As we will demonstrate below for M3, this is orders of magnitude lower than the acceleration caused by the cluster potential, thus it can be neglected.
$a_{g}$ is the acceleration due to the Galactic potential.
Following \citet{2017ApJ...845..148P} it can be written,
\begin{eqnarray}
a_{g} \cdot \vec{n} = -{\rm cos}(b)(\frac{\Theta_{0}^{2}}{R_{0}})({\rm cos}(l) + \frac{\beta}{{\rm sin}^{2}(l) + \beta^{2}}){\rm ~m~s^{-2}},
\end{eqnarray}
where $R_{0}$ is the distance of the Sun from the Galactic center (8.34$\pm$0.16~kpc) and $\Theta_{0}$ is the rotational speed of the Galaxy at the Suns location (240$\pm$8~km~s$^{-1}$)  \citep{2014ApJ...793...51S}.
$\beta = (d/R_{0}){\rm cos}(b) - {\rm cos}(l)$, where $l$ and $b$ are the Galactic longitude and latitude of the GC, respectively. For M3, $l=42.22^{\circ}$ and $b=78.71^{\circ}$) giving $a_{g}\sim -1.201\times10^{-12}\,{\rm m\,s^{-2}}$ .
We notice that compared with the pulsar acceleration observed in M3 (for example, $c(\frac{\dot{P}}{P})_{{\rm obs}}$ $\thicksim$ -5.478 $\times$ 10$^{-11}$~m~s$^{-2}$ for PSR M3E), this effect is negligible, thus, we also ignore it.
$a_{\ell,{\rm GC}}$ is the main contributive term \citep{2005ApJ...621..959F} given by
\begin{eqnarray}
a_{\ell,{\rm GC}}(x) = \frac{9v^{2}}{d\theta_{c}}\frac{\ell}{x^{3}}(\frac{x}{\sqrt{1 + x^{2}}} - {\rm \sinh^{-1}}x),
\end{eqnarray}
where $x$ is the distance from the pulsar to the center of the GC divided by its core radius ($r_{c}$ = $\theta_{c}d$),
$\theta_{c}$ is the core radius (0.37'  for M3),
$v$ is the central stellar velocity dispersion,
and $\ell$ is the distance (also in core radii) to the plane of the sky that passes through the center of the GC.
For the pulsars in M3, we can derive the upper limit of acceleration from $\dot{P}_{{\rm obs}}$ (assuming $\dot{P}_{{\rm int}}$ = 0) \citep{2023ApJ...951L..37L}
\begin{eqnarray}
a_{\ell,{\rm P},{\rm max}} = c(\frac{\dot{P}}{P})_{{\rm obs}},
\end{eqnarray}

\begin{figure*}
\begin{center}
\includegraphics[width=0.721\linewidth]{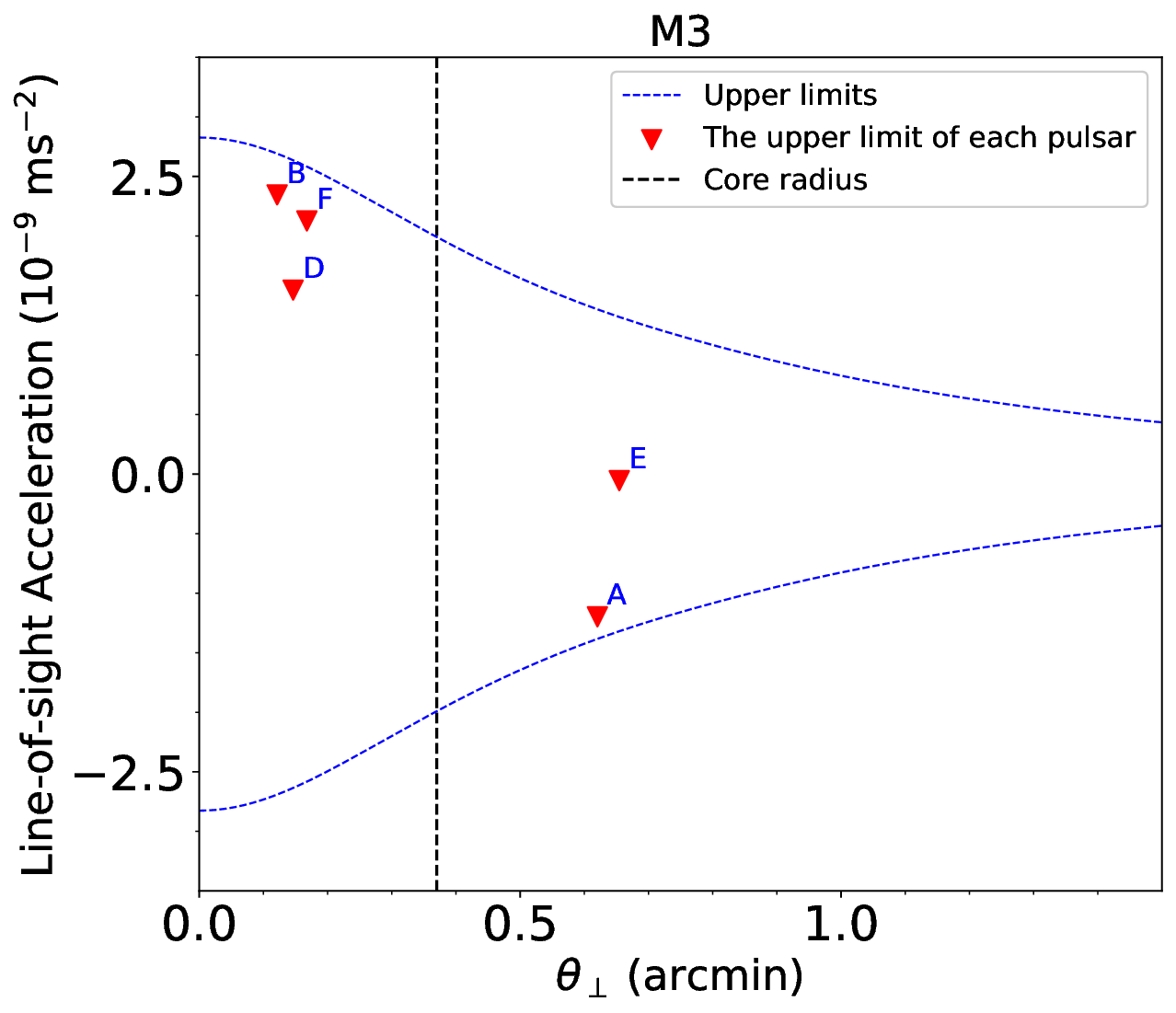}
\caption{Acceleration model for M3. The blue dashed line represents the upper limits for the line-of-sight
pulsar acceleration caused by cluster potential ($a_{\ell,{\rm GC}}$). The abscissa is the constant angular offset from the center ($\theta_{\bot}$). The upper limit of acceleration for each pulsar from the observed spin period derivative was marked with the red triangle. The vertical dashed line represents the core radius (0.37$^{'}$).}\label{fig:4}
\end{center}
\end{figure*}

 In Figure \ref{fig:4}, the blue dashed line represents the range of the maximum acceleration caused by cluster potential $a_{\ell,{\rm GC}}(x)$ along the radius of M3. The upper limit of acceleration for each pulsar from the observed spin period derivative is marked with a red triangle,
 and in the range of maximum acceleration of caused by the cluster potential.
The spin period derivatives of PSRs M3A, E, and F were obtained.
PSR M3A is close to the lower boundary of the range of the maximum acceleration due to the cluster potential. The real distance of M3A to the center can be larger than the projected angular separation. If so, it is possible that the real acceleration of M3A would be larger than the maximum.

\subsection{The Scintillation}\label{sec:scintillation}
The scintillation bandwidth $\Delta\nu_{\rm d}$ and the scintillation timescale $\Delta\tau_{\rm d}$
can be used to quantify the average scintillation characteristics of each observation.
Following \citet{2019MNRAS.485.4389R},
we obtained $\Delta\tau_{\rm d}$ by using a least-square fit to the one-dimensional time-domain ACF,
\begin{eqnarray}
{\rm C}(\Delta\tau,0) = {\rm A\ exp} (- |\frac{\Delta\tau}{\Delta\tau_{\rm d}}|^{\frac{5}{3}})\label{eq:8},
\end{eqnarray}
and
\begin{eqnarray}
{\rm C}(0,0) = {\rm A + W},
\end{eqnarray}
where $A$ is constant to be fitted and $W$ is the noise spike.
After obtaining $A$ and $W$,
we fixed $A$ and got $\Delta\nu_{\rm d}$ by doing a least-squares fit to the one-dimensional frequency-domain ACF via
\begin{eqnarray}
{\rm C}(0,\Delta\nu) = {\rm A\ exp} (- |\frac{\Delta\nu}{\Delta\nu_{\rm d}/{\rm ln2}}|),
\end{eqnarray}

Generally, the uncertainty of scintillation parameters comes from the uncertainty of the $\chi^{2}$ fitting process and the statistical error caused by a finite number of scintles in the dynamic spectra \citep{1999ApJS..121..483B},
\begin{eqnarray}
\sigma_r = (\frac{\mathit{f}BT}{\Delta\nu_{\rm d}\Delta\tau_{\rm d}})^{-0.5},
\end{eqnarray}
where $T$ and $B$ are the integration time and bandwidth, respectively.
The filling factor $\mathit{f}$ is the percentage of scintillation area in the dynamic spectrum.
We calculated the percentage of data with more than 1$\sigma$ in the dynamic spectra and chose $\mathit{f}$ = 0.25 in this work.

From \citet{1990MNRAS.244...68R,1999ApJS..121..483B}, the scintillation strength is
\begin{eqnarray}
u \approx \sqrt{\frac{2\nu}{\Delta\nu_{\rm d}}},
\end{eqnarray}
where $\nu$ is the central frequency (1250~MHz).
When $u<$1, it indicates weak scintillation.
When $u>$1, it indicates strong scintillation.

In some situations,
we can detect arcs in the secondary spectrum.
Scintillation arc can be described as \citet{2019MNRAS.485.4389R}
\begin{eqnarray}
f_{\lambda}=\eta f_{t}^{2},
\end{eqnarray}
in which
\begin{eqnarray}
\label{con:inventoryflow}
\eta   =  1.543 \times 10^{7} \frac{D_{\rm kpc}s(1-s)}{(V_{{\rm eff},\perp}\cos\psi)^{2}}, \label{eq:9}
\end{eqnarray}
where $\eta$ is the so-called arc curvature,
$D_{\rm kpc}$ is pulsar distance in~kpc,
$s$ is the relative distance of the scattering screen to the pulsar,
$\psi$ is the angle between the major axis of the anisotropic structure and the effective velocity vector,
and $V_{{\rm eff},\perp}$ is the effective perpendicular velocity in~${\rm km\ s^{-1}}$.

\begin{figure*}
\begin{center}
\includegraphics[width=18cm,height=8cm]{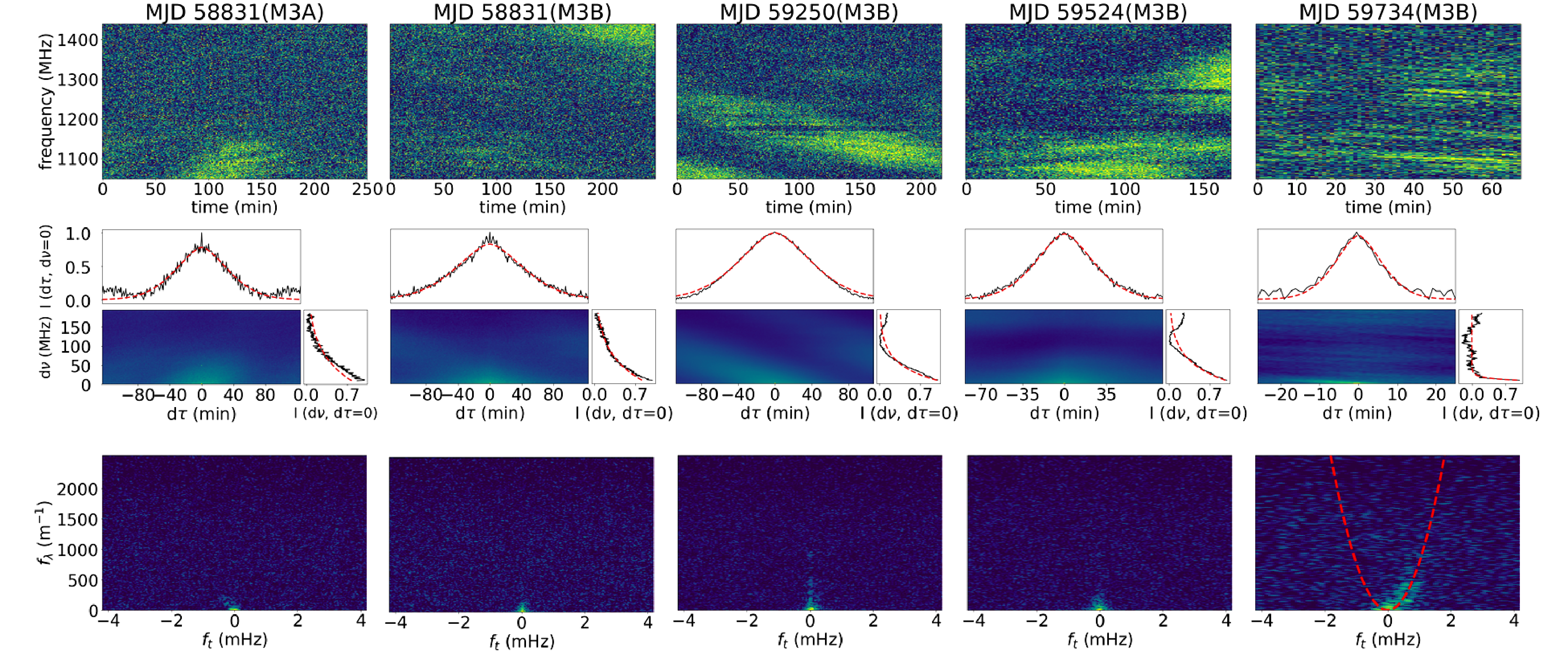}
\caption{Dynamic spectra (top panels), ACFs (middle panels), and secondary spectra (bottom panels) of PSRs~M3A and B. For PSRs~M3A and B, the observations made on MJD 58831, 59250, 59524, 59734 are chosen to show in this figure. The data with more RFI in the first 1200~s are removed. The data duration of PSRs~M3A and B are 249, 249, 216, 167, and 69~min, respectively. In these one-dimensional time-domain and frequency-domain ACFs, the red lines show the best-fit results for $\Delta \tau_d$ and $\Delta \nu_d$, respectively. In all secondary spectra, in order to clearly identify the scintillation arcs, we cut the abscissa around $\pm$4~mHz. In secondary spectrum of PSR~M3B, the red dotted line represent the best-fit arc curvature. }\label{fig:3}
\end{center}
\end{figure*}

In Figure~\ref{fig:3},
we show the dynamic spectra,
ACFs and the secondary spectra for PSRs M3A and B.
We did ACF analysis for all M3 pulsars,
and only obtained the scintillation parameters for PSRs M3A and B as shown in Table~\ref{tab:3}. We remind the reader that as only few patches are captured in most of the dynamic spectra, this may cause additional errors in the measurement of scintillation parameters.

\begin{table}[!ht]
\caption{The scintillation parameters for PSRs M3A and B centered at 1250~MHz.}\label{tab:3}
\centering
\setlength{\tabcolsep}{5.0mm}{
    \centering
    \begin{tabular}{cccccccc}
    \hline\hline
        Pulsar Name & MJD & $A$ &$W$  & $\Delta\tau_{\rm d}$ & $\Delta\nu_{\rm d}$ & $\sigma_{r}$ & $u$ \\
          &  &  &  & (min) & (MHz)&  \\ \hline
        M3A & 58831  & 0.778$\pm$0.013  & 0.222$\pm$0.013 &46.99$\pm$0.99 &57.14$\pm$1.11 & 0.33 & 7  \\
        M3B & 58831  & 0.821$\pm$0.006  & 0.179$\pm$0.006 &60.02$\pm$0.55 &52.99$\pm$0.49  & 0.36 & 7\\
        M3B & 59250  & 0.989$\pm$0.002  & 0.011$\pm$0.002 &55.91$\pm$2.74 &31.84$\pm$0.46 & 0.29 & 9 \\
        M3B & 59524  & 0.973$\pm$0.008  & 0.027$\pm$0.008 &34.43$\pm$2.24 &39.58$\pm$0.80 & 0.29 & 8 \\
        M3B & 59734  & 0.958$\pm$0.042  & 0.042$\pm$0.025 &6.97$\pm$0.26  &4.55 $\pm$0.15  & 0.08 & 23\\\hline
    \end{tabular}}
\leftline{$\textbf{Note}$. $A$ is constant to be fitted in equation \ref{eq:8} and $W$ is the noise spike.  $\Delta\tau_{\rm d}$ and $\Delta\nu_{\rm d}$ are the scintillation timescale and }
\leftline{  bandwidth, respectively. $\sigma_{r}$ is the uncertainty. $u$ is the scintillation strength.}
\end{table}

In the secondary spectra, we only detected a clear scintillation arc for PSR~M3B.
Following \citet{2019MNRAS.485.4389R}, the arc curvature is 649$\pm$23~m${^{-1}}$~mHz$^{-2}$.
It was noticed that there appears to be the beginnings of an arc in the secondary spectrum of PSR~M3A, however because of the weakness of this signal it might be due to noise fluctuations and will not be discussed here.

For PSRs~M3A and B, we did ACF analysis of the dynamic spectra at the same epoch (MJD 58831), and noticed that their scintillation bandwidths are basically the same. With such result, the scattering of PSRs~M3A and B may be dominated by the same screen. More FAST simultaneous observations are needed to confirm this. For PSR~M3B, we also analyzed ACF of the dynamic spectra at three other epochs (MJDs 59250, 59524 and 59734). From Table~\ref{tab:3}, we found that at the same epoch the scintillation timescale of PSRs~M3A and B is different, and the scintillation timescale of PSR~M3B is decreasing by approximately 53~min from MJD 58831 to 59734. We suggest that the differences in the scintillation timescales of PSRs~M3A and B at the same epoch, as well as the variations in the scintillation timescales of PSR~M3B with epoch, could be caused by combinations of the following reasons.
Firstly, when fewer patches are captured in the dynamic spectra, this results in larger errors in the measurement of the scintillation timescale.
Second, as PSRs M3A and B are in binary systems, even at same epoch PSRs M3A and B possess different orbital velocities and at different epochs the orbital velocity of PSR M3B varies with orbital phase.
Third, for PSR~M3B, compared with MJDs 58831, 59250 and 59524, the dynamic spectrum at MJD 59734 has more narrower patches and the scattering may come from different screen. In all the dynamic spectra of PSRs~M3A and B, the scintillation strength $u$ were 7, 7, 9, 8 and 23, respectively, indicating strong scintillation.

The tilted or asymmetrical patterns in the ACF diagrams of PSRs~M3A and B indicate that the corresponding scattering medium is anisotropic.
As described in Section~\ref{sec:Timing result}, PSR~M3A and B are all in binary systems.
To date, as there no systematic drift velocity or orbital velocity measurements for M3 pulsars via VLBI (Very-long-baseline interferometry) or timing,
we could not get the pulsar-to-screen distance from both the measured scintillation timescale and arc curvature by using Equation~\ref{eq:9}.
For pulsars in binary system, multiple observations made at different orbital phases can help us to constrain orbital parameters through the study of the variations of scintillation timescale and arc curvature along orbital phases \citep{2019MNRAS.485.4389R}.
With more FAST observations in the future, our next work will be finding the location of the scattering screens and to constrain the orbital parameters for PSRs~M3A and B.

\section{summary} \label{sec:sum}

The conclusions are as follows:

1. The timing solutions of previously known PSRs~M3A, B, D, E, and F were either updated or obtained for the first time.
With a timing baseline of 937~days, the spin frequency derivative, and the more accurate position of PSR~M3A was measured.
The timing solutions of PSRs~M3B and D are consistent with previous results and updated.
PSRs~M3E and F are binaries with low mass companions.
PSR~M3E is in a 7.1-day orbit and with an eccentricity about 1$\times$10$^{-4}$.
The orbital period of the PSR~M3F circular orbit is about 3~days.

2. After searching 41 different observations with acceleration searches,
the PSR~M3C is still missing.

3. Due to the sensitivity of FAST, the detection rates of pulsars are several to ten times higher than those in \citet{2007ApJ...670..363H}, despite the very strong diffractive scintillation seen towards the cluster at these radio frequencies.
Based on the searching results, the detection rates of PSRs~M3A, B, D, E, and F are 58\%, 90\%, 78\%, 51\%, and 80\%, respectively.

4. As predicted by the low single-binary encounter rate, the characteristics of pulsars in M3 seem to be similar to those of millisecond pulsars in Galactic disk.

5. From the dynamic spectra of PSRs~M3A, B, and with the ACF analysis,
the scintillation timescale ranges from 7.0$\pm$0.3~min to 60.0$\pm$0.6~min,
and the scintillation bandwidth ranges from 4.6$\pm$0.2~MHz to 57.1$\pm$1.1~MHz.
Based on the same epoch analysis, the scattering of PSRs~M3A and B may be dominated by the same screen.
The ISM which is dominating the scattering of PSRs~M3A, and B is anisotropic.
All the dynamic spectra of PSRs~M3A, and B suggested strong scintillation.
From the secondary spectra,
only one scintillation arc of PSR~M3B was obtained with a curvature of 649$\pm$23~m${^{-1}}$~mHz$^{-2}$.
Our next work will be on the scattering screens
and the variation of scintillation parameters at different orbital phases.

\acknowledgments
This work is supported by the National Key R$\&$D Program of China, No. 2022YFC2205202, No. 2020SKA0120100,
the International Centre of Supernovae, Yunnan Key Laboratory (No. 202302AN360001 and 202302AN36000104),
the National Natural Science Foundation of China under Grand No. 11703047, 11773041, U2031119, 12173052, 12173053, 12373032, 11963002, 12041304 and 12288102.
Ju-Mei Yao was supported by the National Science Foundation of Xinjiang Uygur Autonomous Region (2022D01D85), the Major Science and Technology Program of Xinjiang Uygur Autonomous Region (2022A03013-2), the Tianchi Talent project, and the CAS Project for Young Scientists in Basic Research (YSBR-063), and the Tianshan talents program (2023TSYCTD0013).
ZP and LQ are supported by the CAS “Light of West China” Program.
ZP and LQ are supported by the Youth Innovation Promotion Association of the Chinese Academy of Sciences (ID nos. 2023064 and 2018075,  Y2022027).
RPE is supported by the Chinese Academy of Sciences President's International Fellowship Initiative, Grant No. 2021FSM0004.
This work is also supported by the International Centre of Supernovae, Yunnan Key Laboratory.
This work made use of the data from FAST (Five-hundred-meter Aperture Spherical radio Telescope).  FAST is a Chinese national mega-science facility, operated by National Astronomical Observatories, Chinese Academy of Sciences.

\bibliography{file}

\end{document}